\documentclass[prl, twocolumn,aps, floatfix, superscriptaddress,10pt]{revtex4-2}
\usepackage{amsmath} 
\usepackage{amssymb}    
\usepackage{graphicx} 
\usepackage{color, verbatim, comment} 
\usepackage{xcolor}

\usepackage{soul}

\begin{document}

\title{
Anomalous correlators, negative frequencies and non-phase-invariant Hamiltonians in random waves
}

\author{A. Villois} 
\affiliation{School of Engineering, Mathematics and Physics, University of East Anglia, Norwich Research Park, Norwich, NR4 7TJ, United Kingdom}
\affiliation{Dipartimento di Fisica, Universit{\`a} degli Studi di Torino - Via P. Giuria 1, 10125 Torino, Italy}

\author{G. Dematteis} 
\affiliation{Dipartimento di Fisica, Universit{\`a} degli Studi di Torino - Via P. Giuria 1, 10125 Torino, Italy}

\author{Y. V. Lvov} 
\affiliation{Department of Mathematical Sciences, Rensselaer Polytechnic
Institute, Troy, NY, USA.}

\author{M. Onorato}
\affiliation{Dipartimento di Fisica, Universit{\`a} degli Studi di Torino - Via P. Giuria 1, 10125 Torino, Italy}
\affiliation{Istituto Nazionale di Fisica Nucleare, INFN, Sezione di Torino - Via P. Giuria 1, 10125 Torino, Italy}

\author{J. Shatah} 
\affiliation{Courant Institute of Mathematical Sciences, New York University, 251 Mercer St., New York, NY 10012, USA}

\begin{abstract}
We investigate a generic non-phase invariant Hamiltonian model that governs the dynamics of nonlinear dispersive waves. 
We give evidence that initial  data characterized by random phases naturally evolve into phase correlations between positive and negative wavenumbers, leading to the emergence of non-zero anomalous correlators and negative frequencies. Using analytical techniques, we show that anomalous correlators develop on a timescale of $\mathcal{O}(1/\epsilon)$, earlier than the kinetic timescale. Our theoretical predictions are validated through direct numerical simulations of the deterministic system.

\end{abstract}

\maketitle

{\it Introduction. } 
Many physical systems are characterized by the propagation of dispersive and nonlinear waves. When the number of waves is large, a deterministic description becomes impractical, necessitating a statistical approach—similar to the statistical treatment of gases. In the context of interacting waves, since the late 1960s, Wave Turbulence (WT) theory \cite{hasselmann62,Zakharov:67b} has been proposed as an analogue to the Boltzmann theory for gases. A central role in WT theory is played by the Wave Kinetic Equation (WKE), which describes both equilibrium thermodynamic states and out-of-equilibrium dynamics \cite{Nazarenko:11,falkovich1992kolmogorov}.

Despite some rigorous mathematical derivations of the WKE for specific PDEs \cite{deng2023full,staffilani2021wave,deng2023long}, the applicability of WT theory remains an open issue for many PDEs describing nonlinear interacting waves. One key property that has to be proven in the derivation of the WKE is that the Fourier phases of the normal variables remain uncorrelated up to kinetic timescales—this is analogous to the problem of propagation of chaos, a concept also present in the derivation of the Boltzmann equation for hard spheres. Only for the Nonlinear Schrödinger (NLS) equation in dimension $d \ge 2$ this problem  has been rigorously addressed \cite{deng2021propagation}. 

When the nonlinearity is sufficiently strong, phase correlations are expected. This phenomenon is well known in integrable systems, where phase-coherent structures such as solitons or breathers can emerge  and persist even from initial data  characterized by random phases. For nonintegrable systems, the situation is more subtle: in the limit of small nonlinearity, it is not clear if or when phase correlations will appear, and if they will survive for large times. Such correlations can modify the statistical properties of the system, potentially invalidating the assumptions behind the kinetic description, particularly when these correlations emerge from initially random phases.

A special type of phase correlation that can arise is the so-called anomalous correlator. 
These correlators, which first appeared in the BCS theory of superconductivity \cite{BCSTheory}, have been studied in $S$-theory \cite{zakharov1975spin,l2012wave} and are linked to coherent pumping in the system.
 More recently, anomalous correlators have been observed in nonlinear particle chains,  specifically in numerical simulations  of the $\beta$-FPUT system \cite{zaleski2020anomalous}. However, the origin of these correlators in nonlinear dispersive PDEs remains unclear, and a general theoretical understanding is still lacking.

In this Letter, we show that the emergence of anomalous correlators in Hamiltonian wave systems, characterized by four-wave interaction, is associated with the presence of non-phase-preserving terms;  more specifically, we show that terms of the type $3\leftrightarrow1$ are responsible for the phase correlations that ultimately lead to the formation of anomalous correlators. 

{\it Definition of the model.}  We focus our analysis of anomalous correlators on Nonlinear Schr\"odinger  (NLS) equation-like systems.
The standard NLS equation describes dispersive wave systems with four-wave interactions of the type two waves scattering into two waves, which conserve the total number of waves (or wave action). Since conserved quantities in PDEs correspond to symmetries in the system, we modify the NLS equation by introducing terms that break the phase symmetry, thereby violating wave action conservation. However, the Hamiltonian (system's energy) remains conserved despite these modifications.
The Hamiltonian for this system is given by:
\begin{equation}\label{eq:H31}
\begin{aligned}
    &H=H_{\rm lin} + H_{\rm nlin},\\
    &H_{\rm lin}=\int||\partial_x|^{\alpha/2} u|^2dx,\\
    &H_{\rm nlin}=\epsilon \int\left(\frac{\nu}{2} |u|^4+\mu(u^3u^*+(u^*)^3 u)\right)dx,
    \end{aligned}
\end{equation}
where $ u = u(x,t) \in \mathbb{C}$, with \(x\in[0,L]\), and $\alpha$, $\nu$, and $\mu$ are real and positive constants, with $\epsilon \ll 1$ serving as the parameter that scales the nonlinearity relative to dispersion. The parameter $\alpha$ sets the linear wave dispersion relation, see \cite{MMT2}. 
The additional terms involving $\mu$ are responsible for breaking phase invariance. 
The model reduces to the standard NLS equation when $\alpha=2$ and $\mu=0$.
The freedom given by the parameter $\alpha$ is important for the goals of our research: it allows us, for example, to include the dispersion relation of surface gravity waves ($\alpha=1/2$). 
A lot of work is being undertaken on this  application \cite{deng2022wave,berti2024hamiltonian,baldi2018time,korotkevich2024non}, for which the preservation in time of random phases has not yet been proven.

The WT theory deals with the interaction of a large number of normal modes that, in the case of periodic boundary conditions, correspond to the Fourier modes.  Using the following definition of the Fourier transform,
\begin{equation}
    a_k(t)=\frac{1}{L^{1/2}}\int_0^{L} u(x,t)e^{-i\frac{2\pi}{L} kx} dx,
\end{equation}
the equations of motion associated to the Hamiltonian~(\ref{eq:H31}) are an infinite set of coupled ODEs:
\begin{equation}\label{eq:3}
\begin{aligned}
i\frac{d a_1}{d t}&=\omega_1 a_1+
\frac{\epsilon}{L} \sum_{2,3,4} \bigg[\nu a_2^*a_3a_4 \delta_{12}^{34}+\\&\quad+\mu\bigg(a_2a_3a_4\delta_1^{234}+3 a_2^*a_3^*a_4\delta_{123}^4)\bigg)\bigg],
\end{aligned}
\end{equation}
where $\omega_k=|k|^{\alpha}$,  $a_i=a_{k_i}(t)\in \mathbb{C} $ with $a_k\ne a_{-k}^*$,  $\delta_{12...}^{34...}\equiv \delta_{k_1+k_2+...,k_3+k_4+...}$ is the Kronecker delta and the sums over the indices $k_2$, $k_3$ and $k_4$ extend from $\infty$ to $\infty$. 

For our theoretical approach,  it is useful to isolate in the sums the  diagonal and anti-diagonal terms, i.e., those terms for which $k_1=k_3$ and $k_2=k_4$,  $k_1=k_4$ and $k_2=k_4$ (diagonal terms), $k_1=-k_2$ and $k_3=-k_4$ (anti-diagonal terms). The equations of motion now read:

\begin{equation}\label{eq:diag}
\begin{aligned}
i\frac{d a_1}{d t}&=\tilde\omega_1 a_1+ \rho_1 a_{-1}^*+
\frac{\epsilon}{L} \sum^{'}_{2,3,4} \bigg[\nu a_2^*a_3a_4 \delta_{12}^{34}+\\&\quad+\mu\bigg(a_2a_3a_4\delta_1^{234}+3 a_2^*a_3^*a_4\delta_{123}^4)\bigg)\bigg],
\end{aligned}
\end{equation}
where 
\begin{equation}
\label{omega_rho}
\begin{split} 
&
\tilde\omega_1=\omega_1+ \frac{\epsilon}{L} \sum_2  \{(2\nu|a_2|^2+6 \mu \Re [a_2a_{-2}]\},\\
&
\rho_1=\frac{\epsilon}{L}\sum_2  (6\mu|a_2|^2+\nu a_2 a_{-2}),
\end{split}
\end{equation}
and the sums (with a prime) are intended over all terms except for the diagonal and anti-diagonal ones. 
Interestingly, a nonlinear correction to the frequencies, beyond the standard one, is arising from the non-phase invariant contribution. 
In standard WT theory, the main observable is the wave action spectral density function $n_k\equiv n(k,t)$, defined as $n_k=\langle a_k a_k^*\rangle $ (in the limit of $L\rightarrow \infty$), where the average is taken over an initial set of data with random phases and amplitudes~\cite{nazarenko2011wave}, and the time evolution of $n_k$ is described by the WKE, \cite{nazarenko2011wave}. We underline that $\langle a_k a_k^*\rangle$  is not the only existing second-order correlator: one can also build $\langle a_{k_1} a_{k_2}\rangle$. In general, this correlator may appear in non-homogeneous conditions. However, if one assumes statistical homogeneity, the only correlation allowed is for $k_1=-k_2$, see proof in Appendix {B}. This leads to the definition of the so-called {anomalous correlator} as $m_k\equiv m(k,t)=\langle a_k a_{-k}\rangle $~\cite{l2012wave}—note that $m_k\in \mathbb{C} $. \\

{\it Numerical evidence. } 
\begin{figure*}
    \centering
\includegraphics[width=1\textwidth]{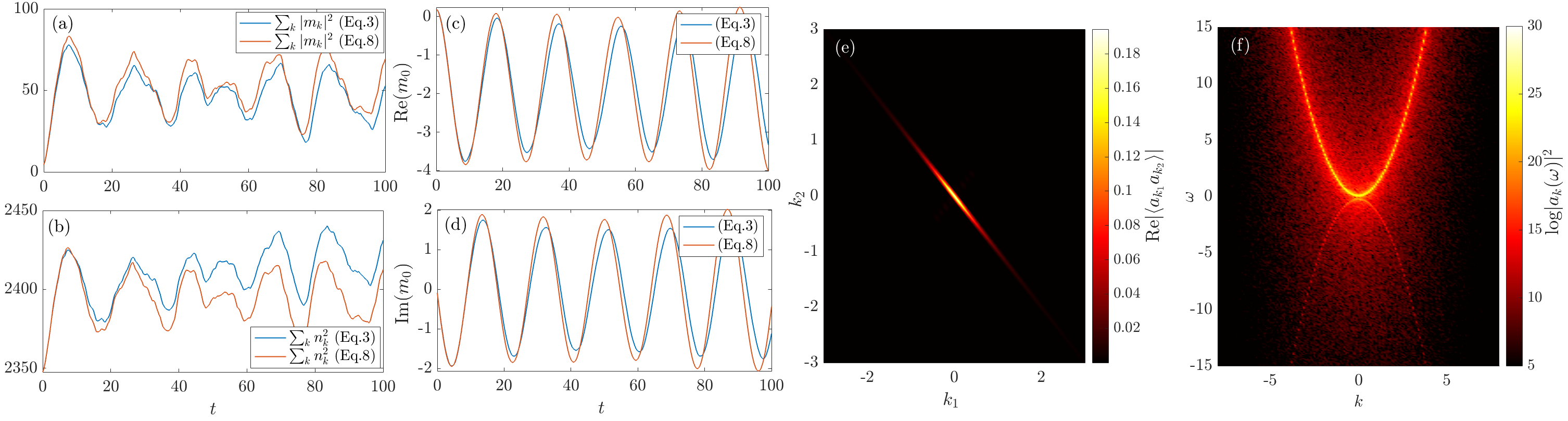}
\caption{(a) Time evolution of the anomalous mass \( \sum_k |m_k|^2 \), showing its rapid growth from an initially zero state. (b) Evolution of the standard correlator \( \sum_k n_k^2 \), which exhibits similar behavior to the anomalous mass. (c) and (d) Evolution of the real and imaginary parts of \( m_k \) for \( k = 0 \), with the real part oscillating around a nonzero value and the imaginary part around zero. 
On these Figures (from (a) to (d)), the blue lines are the result of the ensemble average of numerical simulations of Eq.~\eqref{eq:diag}; the orange line are obtained from the numerical simulations of Eq.~\eqref{eq:5}. The agreement is excellent, even for time scales larger than $\tau=1/\epsilon\sim 11$.
(e) Real part of the correlation matrix $\langle a_{k_1} a_{k_2} \rangle $, obtained by ensemble and time averaging, showing a strong signal along the anti-diagonal, indicating the emergence of anomalous correlations. 
(f) Magnitude of the space-time Fourier transform of $u(x,t)$ over the time window $t = [0,100]$, 
revealing both a positive frequency branch associated with the linear dispersion relation and an additional branch corresponding 
to negative frequencies.
}
    \label{fig:1}
\end{figure*}
Here, we present numerical evidence for the emergence of anomalous correlations in a non-phase-invariant system. Our direct numerical simulations (DNS) integrate Eq.~\eqref{eq:diag} over time, starting from initial conditions where Fourier modes have random, independent, and identically distributed (i.i.d.) uniform phases. The initial mode amplitudes satisfy \( |a_k|^2 = \bar{n}_k \), with the initial smooth distribution given by \( \bar{n}_k = C e^{-k^2 / (2 \sigma^2)} \), where \( \sigma^2 = 40 \). The constant \( C \) is chosen to ensure that the field amplitude in physical space remains \( \mathcal{O}(1) \), enforcing \( \int |u(x,0)|^2 \, dx = L \). We choose the box size to be \( L = 20\pi \). More details on the numerical implementation can be found in Appendix~A.

For the numerical results presented here, we set \( \epsilon = L^{p} \) with \( p = -0.6 \) \cite{buckmaster2021onset}, along with \( \nu = 1 \), \( \mu = 0.1 \), corresponding to a weakly nonlinear regime where the ratio of linear to nonlinear contributions satisfies \( H_{\rm lin} / H_{\rm nlin} \simeq \mathcal{O}(10^2) \). The choice of $\mu$ fulfills the stability condition $\mu<\nu/3$ (see Appendix D). 
Simulations were performed for various values of \( \alpha \), including \( \alpha = 1/2 \), all yielding qualitatively similar results. The results shown here correspond to \( \alpha = 2 \); 
 for such value of $\alpha$, resonant interactions of the type $3\leftrightarrow 1$ are present, i.e., there exist four wavenumbers such that the two equations below are satisfied simultaneously:
\begin{equation}
k_1=k_2+k_3+k_4,\;\ k_1^2=k_2^2+k_3^2+k_4^2.
\end{equation}
Figure~\ref{fig:1} illustrates the development of anomalous correlations  from random initial data. 
Fig.\ref{fig:1}(a) 
displays the evolution of the anomalous mass, defined as \( \sum_k |m_k|^2 \). The blue curves represent ensemble averages computed over 500 realizations of the initial ensemble. Initially zero, the anomalous mass rapidly grows to nonzero values. For comparison, Fig.\ref{fig:1}(b) shows the evolution of the wave action \( n_k \), where \( \sum_k n_k^2 \) exhibits behavior similar to that of the anomalous mass.

Focusing on the time evolution individual modes, Fig.\ref{fig:1}(c) shows the real part of \( m_k \) for \( k = 0 \), which oscillates around a nonzero value. 
To provide a more complete picture of the development of anomalous correlations, Fig.\ref{fig:1}(e) of Fig.~\ref{fig:1} shows the real part of the correlation matrix \( \langle a_{k_1} a_{k_2} \rangle \). This matrix is obtained by averaging over different realizations and subsequently time-averaging over the simulation window. A clear signal appears along the anti-diagonal, indicating the spontaneous emergence of anomalous correlations. Notably, for the phase-invariant case (\( \mu = 0 \)), the same simulations show no such anti-diagonal correlations, and the anomalous mass remains zero throughout the entire evolution.

Finally, we analyze the system in both space and time by performing a Fourier transform over the time window \( t = [0,100] \). Fig.\ref{fig:1}(f) shows the magnitude of the space-time Fourier transform, revealing not only the expected positive frequency branch associated with the linear dispersion relation, but also an additional branch corresponding to negative frequencies. The presence of negative frequencies and their connection to anomalous correlators in non-phase-invariant Hamiltonian systems will be further explored in subsequent sections.

{\it Theoretical developments. }
To provide theoretical understanding of the numerical results, we start by stating the following {\bf Theorem}:
{\it A phase-invariant Hamiltonian system with initial data distributed according to a phase-invariant measure  cannot develop anomalous correlations.}
We give a proof of the theorem in Appendix B. Examples of phase-invariant measures include the gaussian  or the random-phase measure (adopted in our numerical simulations). 
Note that this theorem is valid for any level of nonlinearity, i.e. it is not restricted to the weakly nonlinear regime.

Now, starting from equations \eqref{eq:diag}, the goal is to derive evolution equations for the observables $n_k(t)$ and $m_k(t)$ previously defined; we will  show that, if $m_k(t=0)=0$, then the anomalous correlators will arise at a time scale of order $1/\epsilon$ (we recall that the time scale of the standard WKE, when a kinetic regime exists, is $1/\epsilon^2$, \cite{nazarenko2011wave,deng2023full}. 
A key step in the derivation is the generalization of the Wick selection rule by which the fourth-order correlators are split by including the anomalous correlators \cite{LVOVBOOK}:
\begin{equation}\label{eq:Wick}
\begin{split}
&\langle a_1^*a_2^*a_3 a_4\rangle=n_1 n_2(\delta_1^3\delta_2^4+\delta_1^4\delta_2^3)+m_1^*m_3
\delta_{12}\delta_{34}\,,\\
&\langle a_1^* a_2a_3 a_4\rangle=
n_1 m_3(\delta_1^2\delta_{34}+
\delta_1^4\delta_{23})+
n_1 m_4\delta_1^3\delta_{24}\,,\\
&\langle a_1 a_2a_3 a_4\rangle=
m_1 m_3(\delta_{12}\delta_{34}+
\delta_{14}\delta_{23})+
m_1 m_4\delta_{13}\delta_{24}\,.
\end{split}
\end{equation}
Assuming Random Phase and Amplitude (RPA) initial data \cite{nazarenko2011wave}, the resulting equations at order $\epsilon$ are the following (details in Appendix C):
\begin{equation}\label{eq:5}
\begin{split}
&\frac{d n_k}{dt}=
2\Im[\langle \rho_k \rangle m_k^*],
\\ &
i \frac{d m_k}{dt}=2\langle\tilde \omega_k\rangle m_k +\langle \rho_k \rangle ( n_k+ n_{-k}),
\end{split}
\end{equation}
where 
\begin{equation}
\begin{split}
&
\langle \tilde\omega_k\rangle=\omega_k+ \frac{\epsilon}{L} \sum_2 \{(2\nu n_2+6 \mu \Re [m_2]\},\\
&
\langle\rho_k\rangle=\frac{\epsilon}{L}\sum_2 (6\mu  n_2+\nu m_2).
\end{split}
\end{equation}
Note that after some algebra manipulation and the assumption of isotropy in the wave action, $n_k=n_{-k}$, it is possible to prove the following result
\begin{equation}\label{eq.derivatives}
    \frac{d}{dt} n_k^2=\frac{d}{dt} |m_k|^2.
\end{equation}
In Fig.~\ref{fig:1}, panels (a)–(b), the orange lines represent the time evolution of Eqs.~\eqref{eq:5}, initialized with the same conditions as the ensemble averages obtained from Eq.~\eqref{eq:diag}. This shows that Eqs.~\eqref{eq:5} accurately capture the evolution of \( n_k \) and \( m_k \) over the characteristic timescale \( \mathcal{O}(1/\epsilon) \). Moreover, the similarity in behavior observed in both figures is justified by the relation given in Eq.~\eqref{eq.derivatives}.  
Under the RPA assumption, we have \( {m}_k(t=0) = 0 \) while \( {n}_k(t=0) \neq 0 \). Consequently, if \( \mu = 0 \), anomalous correlators cannot develop, in agreement with Theorem previously stated. However, at order \( \epsilon \), anomalous correlators emerge only if the system lacks phase invariance. This effect arises due to a driving term in the evolution equation for \( m_k \), specifically the term \( \epsilon \sum_2 (6 \mu n_2)(n_k + n_{-k}) \) in Eq.~\eqref{eq:5}. This term acts as a forcing mechanism on the real part of \( m_k \), solely dependent on \( \mu \) and on the standard correlators \( n_k \). The presence of this driving term explains the offset in the oscillations of the anomalous correlator seen in Fig.\ref{fig:1} (c).


{\it  Negative frequencies and anomalous correlators: }
\begin{figure}
\centering
\includegraphics[width=0.9\columnwidth]{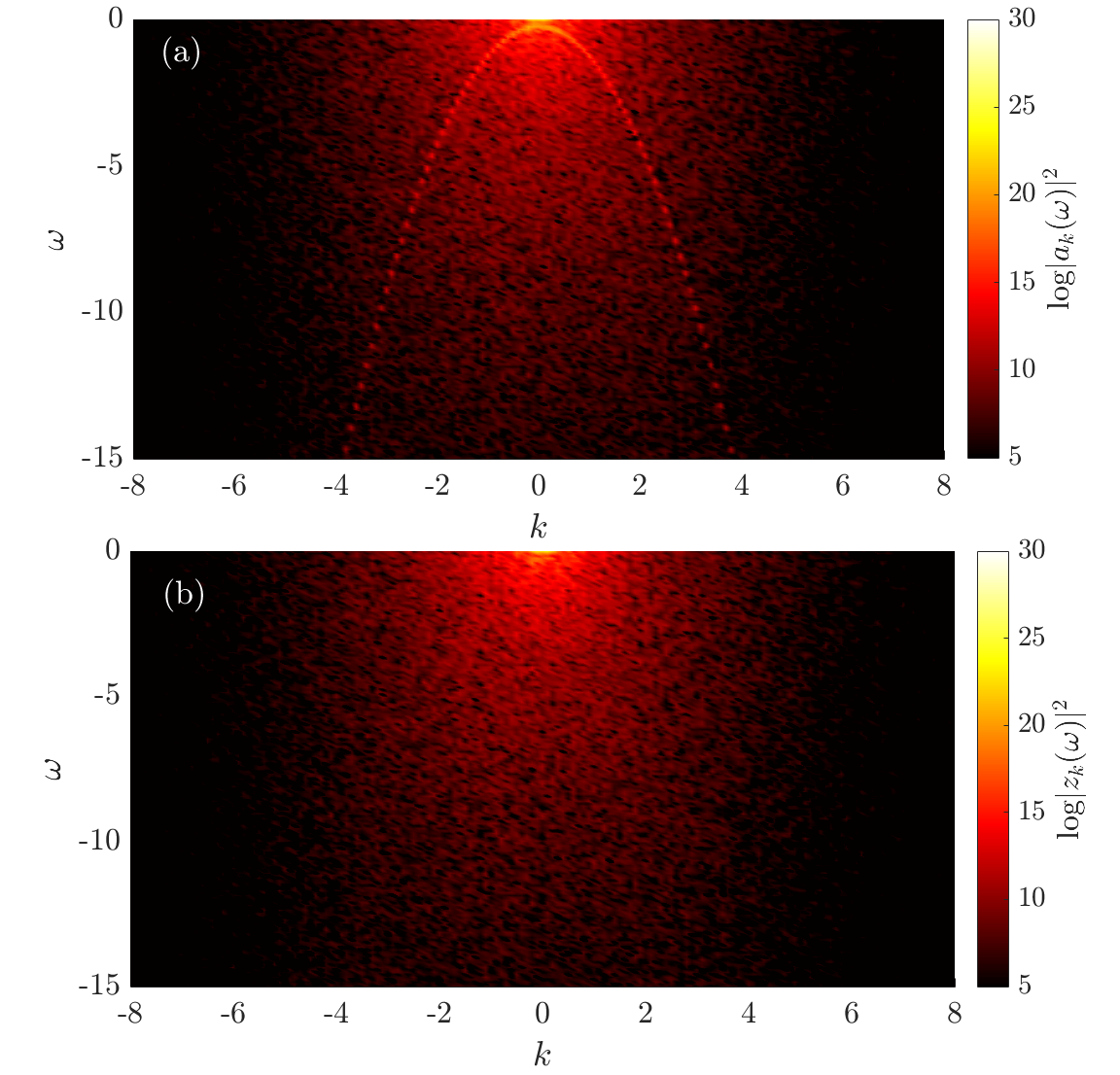}
\caption{The figure illustrates the dispersion relations obtained through spatio-temporal Fourier transform for a single realization of the dynamics shown in Fig.~\ref{fig:1}. (a) shows the negative frequencies branch of dispersion relation obtained using variables $a_k$. (b) shows the absence of a negative frequency branch when variables $z_k$ are used.}
\label{fig:2}
\end{figure}
The set of equations in (\ref{eq:3}) appear rather complicated; for interpretation purposes, we focus our attention on a {\it minimal model}, i.e., a simplified version of our equations that still retains sufficient information that allows us to establish a link between the negative frequencies observed numerically and the existence of anomalous correlators. We propose the following  linear set of ODEs:
\begin{equation}\label{eq:7}
i\frac{d a_k}{d t}=\bar \omega_k a_k+ \bar\rho a_{-k}^*,
\end{equation}
with 
\begin{equation}
\bar\omega_k=\omega_k+2\epsilon\nu A,\;\;\;
\bar\rho=6\epsilon \mu A
\end{equation}
with $A=\frac{1}{L}\sum_l|a_l(t=0)|^2$. In our {\it minimal} model, we have simplified the original equations in (\ref{eq:diag})–(\ref{omega_rho}) by neglecting the sums in (\ref{eq:diag}) and the terms in (\ref{omega_rho}) proportional to $a_ka_{-k}$. This last approximation is based on the assumption that these terms, which are associated with the anomalous correlator, are small. Moreover, we have taken $A$ to be time-independent (fixed by the initial conditions), assuming that it evolves on a slower time scale with respect to the other terms. Although these assumptions are rather crude and not fully justified, they yield reasonable results (also in the framework of the full model), as will be shown later.
The linear system above is diagonalized by the transformation from $a_k$ to $z_k$
\begin{equation}\label{eq:8}
a_k= z_k - c_k z_{-k}^*\,,\quad a_{-k}^*= z_{-k}^* - c_k z_{k}\,,
\end{equation}
with 
\begin{equation}
\label{eq:lambda_c}
 c_k= \frac{\bar\rho}{\lambda_k + \bar\omega_k},\,\,\lambda_k= \sqrt{\tilde\omega_k^2 - \bar\rho^2}.
 \end{equation}
 We provide the analytical details of the diagonalization of Eq.~\eqref{eq:7} in Appendix D.
The evolution equations for $z_k$ and $z_{-k}^*$ are decoupled:
\begin{equation}
i\frac{d z_k}{d t}=\lambda_k z_k\,,\quad i\frac{d z_{-1}^*}{d t}=\lambda_kz_{-k}^*\,, 
\end{equation}
with solution
\begin{equation}\label{eq:9}
z_k(t)=z_k(0)e^{-i\lambda_k t}\,,\quad z_{-k}^*(t)=z_{-k}^*(0)e^{i\lambda_k t}\,,
\end{equation}
 which leads to the solution for $a_k$
\begin{equation}\label{eq:sol_a}
a_k(t)=z_k(0)e^{-i\lambda_k t}-c_k z_{-k}^*(0)e^{i\lambda_k t}.
\end{equation}
The $a_k$ variables oscillate with two different frequencies, $\pm\lambda_k$, one of which is responsible for the negative branch displayed in  Fig.\ref{fig:1}(f). Note also that these two branches exhibit different intensities due to the presence of two distinct coefficients in Eq. (\ref{eq:sol_a}).
Moreover, it is interesting to note that if there are no correlations at time $t=0$, the normal variables $z_k$ do not develop them. Conversely, if the anomalous correlator is nonzero at $t=0$, it will persist indefinitely.

The diagonalization provides an explanation to a simple mechanism of formation of anomalous correlation rooted in the linear dynamical coupling of the conjugate variable pair $\{a_k,a_{-k}^*\}$ in Eq.~\eqref{eq:7}. 
Indeed, posing that
\begin{equation}
\langle z_k(0) z_{k}(0)^* \rangle = n_k^{(z)}(0)\,,\quad \langle z_k(0) z_{-k}(0) \rangle  =0\,,
\end{equation}\label{eq:corr_zk}
using (\ref{eq:sol_a}), we have:
\begin{equation}\label{eq:12}
\begin{aligned}
m_k(t) = & \langle a_k(t) a_{-k}(t) \rangle = -c_k \Big(n_k^{(z)}(0) + n_{-k}^{(z)}(0)\Big) 
\end{aligned}
\end{equation}
i.e., an anomalous correlation is created spontaneously for $a_k$, while it will never exist for the normal variable $z_k$, for which
\begin{equation}\label{eq:corr_zk}
    m_k^{(z)}(t) = \langle z_k(t) z_{-k}(t) \rangle =m_k^{(z)}(0) = 0\,. 
\end{equation}

\begin{figure}
    \centering  \includegraphics[width=0.9\columnwidth]{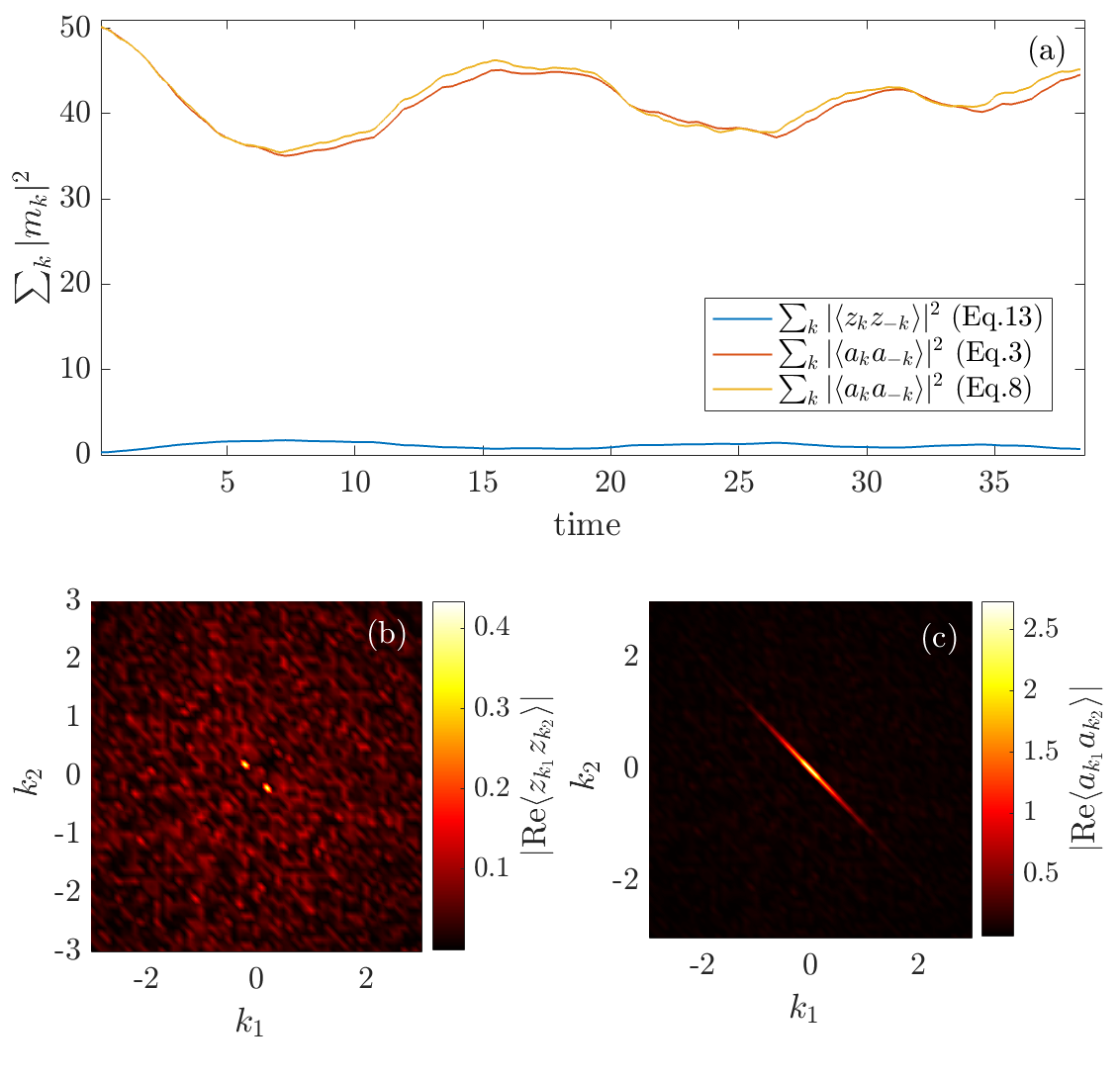}
   \caption{
The figure shows the evolution of the anomalous mass evaluated for the variables $z_k$ and $a_k$, represented by the blue and red lines, respectively. The yellow line indicates the anomalous mass predicted by Eq.~\eqref{eq:5}. Figures (b) and (c) display the correlation matrix evaluated at the final time for the variables $z_k$ and $a_k$ respectively.}    \label{fig:3}
\end{figure}

{\it Further numerical validation. } The analytical results presented so far are further supported by numerical simulations, as shown in Figs.~\ref{fig:2}–\ref{fig:3}. 
Figure~\ref{fig:2}(a) provides a zoomed-in view of the negative frequency branch previously observed in Fig.~\ref{fig:1}(f). In contrast, Fig.~\ref{fig:2}(b) presents the space-time Fourier transform of the action density for the $z_k$ variables computed at each time step using Eqs.~\eqref{eq:8}, which represents the outcome of our {\it minimal} model.
Despite the fact that such a model is not derived by a strict mathematical procedure, the numerical results show that the negative frequency branch has disappeared from the plot. 
To reinforce our findings, we examine the case of initial data that are uncorrelated in the $z_k$ variables. In Fig.~\ref{fig:3}(a), we compare the anomalous mass evaluated using the $z_k$ variables with the one obtained for the $a_k$ variables. As shown by the plot, the anomalous mass in the $z_k$ variables remains nearly zero throughout the time evolution. This result confirms that for uncorrelated initial data in $z_k$, see Fig.~\ref{fig:3}(b), there is no development of anomalous correlators, as predicted by Eq.~\eqref{eq:corr_zk}. On contrary, our chosen initial data exhibit anomalous correlations in the $a_k$ fields, see Fig.~\ref{fig:3}(c). This is not only evident in Fig.~3 but is also accurately captured by Eq.~\eqref{eq:5}.\\
{\it Conclusions. } {\color{black}
In this work, we have identified a general criterion for the emergence of anomalous correlators in wave systems: the presence of phase symmetry-breaking terms in the Hamiltonian. This criterion provides a unifying explanation for the appearance of anomalous correlators in seemingly unrelated systems, such as the FPUT model, BCS theory, and S-theory, all of which exhibit explicit phase symmetry breaking in their Hamiltonians.  
We have provided evidence that anomalous correlators influence the standard correlator \( \langle a_k a_k^* \rangle \) up to a timescale of \( 1/\epsilon \). In addition to the emergence of anomalous correlators \( \langle a_k a_{-k} \rangle \), we showed that non-phase-invariant systems also generate negative frequency modes. To address this, we introduced a normal variable transformation, which eliminates negative frequencies and provides a more effective framework for handling anomalous correlators.
Notably, if normal variables are initially uncorrelated, this uncorrelation is preserved at times up to \( 1/\epsilon \).  
These findings have important implications for wave turbulence theory. Since standard derivations of the WKE rely on the assumption that random phases remain uncorrelated, the spontaneous generation of anomalous correlators presents a fundamental challenge to the validity of the WKE in systems with phase-symmetry breaking. 
Resolving this issue is a crucial step toward a rigorous derivation of kinetic equations in such systems.  
Looking ahead, this work lays the foundation for future research aimed at reconciling kinetic descriptions with the effects of phase symmetry breaking, ultimately leading to a deeper understanding of wave turbulence in non-phase-invariant systems.}\\
{\bf Acknowledgments}
M.O. was funded by Progetti di Ricerca di Interesse Nazionale (PRIN), Project No. 2020X4T57A and 2022WKRYNL,  project “Mathematical Methods in Non-Linear Physics” and “FieldTurb”, Istituto Nazionale di Fisica Nucleare.  The Simons Foundation, United States, Award 652354 on Wave Turbulence is acknowledged for support.\\
\appendix
\section{Appendix A: Numerical details}
Numerically, we solve the following equation of motion:
\begin{equation}\label{eq:3_galerkin}
i \frac{\partial u}{\partial t} = \mathcal{P} \left[ \partial_x^\alpha u + \epsilon \left( \nu \mathcal{P}(|u|^2) + \mu \mathcal{P}((u^*)^2) + 3 \mu \mathcal{P}(u^3) \right) u \right],
\end{equation}
where \(\mathcal{P}\) is the Galerkin projector, which acts in Fourier space by filtering out all \(|k|\)-modes higher than \(|2/3 k_{\text{max}}|\), where $k_{\text{max}} = \pi/\Delta x$,
with $\Delta x$  the resolution in physical space. 
The numerical integration is performed pseudo-spectrally: the linear part of Eq.~\eqref{eq:3_galerkin} is solved in Fourier space, while the nonlinear part is handled using a fourth-order Runge-Kutta method. The time step is chosen to be at least 10 times smaller than the fastest linear oscillation.
\section{Appendix B: Proof of the theorem for phase-invariant systems} Let $\gamma =\{\gamma_k\}$, with $k\in\frac{2\pi}{L}\mathbb Z$,  be complex-valued i.i.d. random variables, and let $f(x,\gamma) = \sum_k \gamma_k \sqrt{\bar n_{k}} e^{ i k x}$ denote a random field. Assume that the measure of $\gamma_k$ ($dF$) is phase-invariant,
\begin{equation}
\langle G (e^{i\alpha}z)\rangle = \langle G (z)\rangle\,,
\end{equation}
where $\langle G(z)\rangle :=\int G( z )dF(z)$. This is the case of $z=z_1+iz_2$ and $dF = \frac{1}{\sqrt{2\pi}}e^{i |z|^2/2}dz_1dz_2$, or $z=e^{i\theta}$ and $dF = d\theta$. If $u(x,t;\gamma)= \sum_k a_k(t;\gamma) e^{i k x}$ solves a translation-invariant equation with initial value $a_k(0;\gamma)=\sqrt{\bar n_{k}}\gamma_k$, then $u(x+\alpha,t;\gamma)= \sum_k a_k(t;\gamma) e^{i k (x+\alpha)}=\sum_k e^{i k x}a_k(t;\gamma) e^{i k \alpha}$ solves the equation with initial data $\sum_k e^{i k x}\sqrt{\bar n_{k}}\gamma_k e^{i k \alpha}$. Consequently, $a_k(t;\gamma_\alpha)=e^{ik\alpha} a_k(t;\gamma)$, with $\gamma_\alpha={e^{ik\alpha}\gamma_k}$. Therefore,
$\langle e^{i(k-l)\alpha}a_k(t;\gamma) a_l^*(t;\gamma)\rangle = \langle a_k(t;\gamma_\alpha) a_l^*(t;\gamma_\alpha)\rangle = \langle a_k(t;\gamma) a_l^*(t;\gamma)\rangle\,,$
where the last equivalence is true thanks to the phase-invariance of the measure. Thus,
\begin{equation}
    \langle a_k(t;\gamma) a_l^*(t;\gamma)\rangle = n_k(t) \delta(k-l)\,.
\end{equation}
Similarly, 
\begin{equation}
    \langle a_k(t;\gamma) a_l(t;\gamma)\rangle = m_k(t) \delta(k+l)\,.
\end{equation}
Note that, since $\{\gamma_k\}$ are i.i.d., $m_k(0)=0$, because $\langle \gamma_k \gamma_{-k} \rangle=0$. 
If the evolution is phase-invariant, i.e. 
\begin{equation}
    a_k(t;e^{i\alpha}\gamma) = e^{i\alpha}a_k(t;\gamma)\,, 
\end{equation}
then
\begin{equation}
\begin{aligned}
    e^{2i\alpha} m_k(t)  &= e^{2i\alpha}\langle a_k(t;\gamma) a_{-k}(t;\gamma)\rangle \\
    &= \langle a_k(t;\gamma_\alpha) a_{-k}(t;\gamma_\alpha)\rangle =m_k(t)\,. 
    \end{aligned}
\end{equation}
Therefore, for $\{\gamma_k\}$ i.i.d., phase-invariant measure and phase-invariant evolution, $m_k(t)=0$ for any $t$.
Note that if $\{\gamma_k\}$ are i.i.d. with phase invariant law, but the evolution is not phase-invariant, even if $m_k(0)=0$, $m_k(t)$ is not necessarily zero for $t>0$. 
\section{Appendix C: Derivation of Eq.~\eqref{eq:5}}
Here, we derive the two Eqs.~\eqref{eq:5} starting from Eq.~\eqref{eq:diag}.
Let us multiply \eqref{eq:diag} by $a_1^*$, and then subtract the complex conjugate of \eqref{eq:diag} multiplied by $a_1$. This yields:
\begin{equation}
    \begin{aligned}
        i\frac{d}{dt}|a_1|^2 & = \tilde \omega_1 (|a_1|^2-|a_1|^2) + \rho_1 (a_1 a_{-1})^* - \rho_1^* a_1 a_{-1}+o.t.\\
        & = 2i \Im [\rho_1 (a_1 a_{-1})^*]+o.t.
    \end{aligned}
\end{equation}
where $o.t.$ indicates the remainder of off-diagonals terms.

Now, let us multiply \eqref{eq:diag} by $a_{-1}$, and then add \eqref{eq:diag} but replacing $k_1$ with $-k_1$, and multiplied by $a_1$. We obtain:
\begin{equation}
    \begin{aligned}
        i\frac{d}{dt}(a_1a_{-1}) & = 2\tilde \omega_1 (a_1 a_{-1}) + \rho_1 |a_{-1}|^2 + \rho_{-1} |a_{1}|^2+o.t.\\
        & = 2i \Im [\rho_1 (a_1 a_{-1})^*]+o.t.
    \end{aligned}
\end{equation}
Averaging over initial data and assuming that the system remains spatially homogeneous for later times, then we obtain Eqs.~\eqref{eq:5}. 

\section{Appendix D: Diagonalization of the linear non-phase-invariant dynamics}
Rewrite the linear system~\eqref{eq:7} in simplified notation as
\begin{equation}\label{eq:7A}
i\frac{d u}{d t}=\bar \omega u+ \bar \rho v\,, \qquad i\frac{d w}{d t}= -\bar\omega v - \bar \rho u\,,
\end{equation}
or in matrix form as
\begin{equation}\label{eq:8A}
i\frac{d }{d t}\left(\begin{array}{c}u\\  v\end{array}\right)=\left(\begin{array}{cc}\bar \omega & \bar \rho\\ -\bar \rho & -\bar \omega \end{array}\right) \left(\begin{array}{c}u\\  v\end{array}\right)\,.
\end{equation}
Here, we use the following short-hand notation: $u=a_k$, $v=a_{-k}^*$, $\bar \rho=6\epsilon\mu A$, $\bar\omega=\omega+2\epsilon\nu A$.
The matrix has eigenvalues $\pm\lambda$, with $\lambda=\sqrt{\bar \omega^2 - \bar \rho^2}$, with eigenvectors given respectively by
\begin{equation}
    e_1 = \left(\begin{array}{c}1\\  -c\end{array}\right)\,,\quad e_2 = \left(\begin{array}{c}-c\\  1\end{array}\right)\,,\quad \text{with } c=\frac{\bar \rho}{\lambda + \bar\omega}\,.
\end{equation}
Notice that the symmetry between $e_1$ and $e_2$ is due to the peculiar relationship between $\bar \rho$, $\lambda$, and $\bar \omega$, by which $(\lambda-\bar \omega)/\bar \rho=-\bar \rho/(\lambda+\bar \omega)=-c$.
The diagonalizing change of variables is thus given by
\begin{equation}\label{eq:10A}
\left(\begin{array}{c}u\\  v\end{array}\right)=\left(\begin{array}{cc}1 & -c\\ -c & 1 \end{array}\right) \left(\begin{array}{c}z\\  w\end{array}\right)\,,
\end{equation}
so that $u=z-cw$ and $z=\frac{1}{1-c^2}(u+cv)$. Finally, notice that for $\lambda$ to be real when $\omega=0$ we need the condition $\mu<\nu/3$. This is a condition for stability that must be satisfied by the Hamiltonian \eqref{eq:H31}.

\newpage

\begin{thebibliography}{20}%
\makeatletter
\providecommand \@ifxundefined [1]{%
 \@ifx{#1\undefined}
}%
\providecommand \@ifnum [1]{%
 \ifnum #1\expandafter \@firstoftwo
 \else \expandafter \@secondoftwo
 \fi
}%
\providecommand \@ifx [1]{%
 \ifx #1\expandafter \@firstoftwo
 \else \expandafter \@secondoftwo
 \fi
}%
\providecommand \natexlab [1]{#1}%
\providecommand \enquote  [1]{``#1''}%
\providecommand \bibnamefont  [1]{#1}%
\providecommand \bibfnamefont [1]{#1}%
\providecommand \citenamefont [1]{#1}%
\providecommand \href@noop [0]{\@secondoftwo}%
\providecommand \href [0]{\begingroup \@sanitize@url \@href}%
\providecommand \@href[1]{\@@startlink{#1}\@@href}%
\providecommand \@@href[1]{\endgroup#1\@@endlink}%
\providecommand \@sanitize@url [0]{\catcode `\\12\catcode `\$12\catcode
  `\&12\catcode `\#12\catcode `\^12\catcode `\_12\catcode `\%12\relax}%
\providecommand \@@startlink[1]{}%
\providecommand \@@endlink[0]{}%
\providecommand \url  [0]{\begingroup\@sanitize@url \@url }%
\providecommand \@url [1]{\endgroup\@href {#1}{\urlprefix }}%
\providecommand \urlprefix  [0]{URL }%
\providecommand \Eprint [0]{\href }%
\providecommand \doibase [0]{https://doi.org/}%
\providecommand \selectlanguage [0]{\@gobble}%
\providecommand \bibinfo  [0]{\@secondoftwo}%
\providecommand \bibfield  [0]{\@secondoftwo}%
\providecommand \translation [1]{[#1]}%
\providecommand \BibitemOpen [0]{}%
\providecommand \bibitemStop [0]{}%
\providecommand \bibitemNoStop [0]{.\EOS\space}%
\providecommand \EOS [0]{\spacefactor3000\relax}%
\providecommand \BibitemShut  [1]{\csname bibitem#1\endcsname}%
\let\auto@bib@innerbib\@empty
\bibitem [{\citenamefont {Hasselmann}(1962)}]{hasselmann62}%
  \BibitemOpen
  \bibfield  {author} {\bibinfo {author} {\bibfnamefont {K.}~\bibnamefont
  {Hasselmann}},\ }\bibfield  {title} {\bibinfo {title} {{On the non--linear
  energy transfer in a gravity--wave spectrum. Part I: General theory}},\
  }\href@noop {} {\bibfield  {journal} {\bibinfo  {journal} {J. Fluid Mech.}\
  }\textbf {\bibinfo {volume} {12}},\ \bibinfo {pages} {481} (\bibinfo {year}
  {1962})}\BibitemShut {NoStop}%
\bibitem [{\citenamefont {Zakharov}\ and\ \citenamefont
  {Filonenko}(1967)}]{Zakharov:67b}%
  \BibitemOpen
  \bibfield  {author} {\bibinfo {author} {\bibfnamefont {V.~E.}\ \bibnamefont
  {Zakharov}}\ and\ \bibinfo {author} {\bibfnamefont {N.~N.}\ \bibnamefont
  {Filonenko}},\ }\href@noop {} {\bibfield  {journal} {\bibinfo  {journal} {J.
  Appl. Mech. Tech. Phys.}\ }\textbf {\bibinfo {volume} {4}},\ \bibinfo {pages}
  {506} (\bibinfo {year} {1967})}\BibitemShut {NoStop}%
\bibitem [{\citenamefont {Nazarenko}(2011{\natexlab{a}})}]{Nazarenko:11}%
  \BibitemOpen
  \bibfield  {author} {\bibinfo {author} {\bibfnamefont {S.}~\bibnamefont
  {Nazarenko}},\ }\href@noop {} {\emph {\bibinfo {title} {Wave Turbulence}}}\
  (\bibinfo  {publisher} {Springer-Verlag},\ \bibinfo {address} {Heidelberg},\
  \bibinfo {year} {2011})\BibitemShut {NoStop}%
\bibitem [{\citenamefont {Falkovich}\ \emph {et~al.}(1992)\citenamefont
  {Falkovich}, \citenamefont {Lvov},\ and\ \citenamefont
  {Zakharov}}]{falkovich1992kolmogorov}%
  \BibitemOpen
  \bibfield  {author} {\bibinfo {author} {\bibfnamefont {G.}~\bibnamefont
  {Falkovich}}, \bibinfo {author} {\bibfnamefont {V.~S.}\ \bibnamefont
  {Lvov}},\ and\ \bibinfo {author} {\bibfnamefont {V.~E.}\ \bibnamefont
  {Zakharov}},\ }\href@noop {} {\emph {\bibinfo {title} {Kolmogorov spectra of
  turbulence}}}\ (\bibinfo  {publisher} {Springer, Berlin},\ \bibinfo {year}
  {1992})\BibitemShut {NoStop}%
\bibitem [{\citenamefont {Deng}\ and\ \citenamefont
  {Hani}(2023{\natexlab{a}})}]{deng2023full}%
  \BibitemOpen
  \bibfield  {author} {\bibinfo {author} {\bibfnamefont {Y.}~\bibnamefont
  {Deng}}\ and\ \bibinfo {author} {\bibfnamefont {Z.}~\bibnamefont {Hani}},\
  }\bibfield  {title} {\bibinfo {title} {Full derivation of the wave kinetic
  equation},\ }\href@noop {} {\bibfield  {journal} {\bibinfo  {journal}
  {Inventiones mathematicae}\ ,\ \bibinfo {pages} {1}} (\bibinfo {year}
  {2023}{\natexlab{a}})}\BibitemShut {NoStop}%
\bibitem [{\citenamefont {Staffilani}\ and\ \citenamefont
  {Tran}(2021)}]{staffilani2021wave}%
  \BibitemOpen
  \bibfield  {author} {\bibinfo {author} {\bibfnamefont {G.}~\bibnamefont
  {Staffilani}}\ and\ \bibinfo {author} {\bibfnamefont {M.-B.}\ \bibnamefont
  {Tran}},\ }\bibfield  {title} {\bibinfo {title} {On the wave turbulence
  theory for stochastic and random multidimensional kdv type equations},\
  }\href@noop {} {\bibfield  {journal} {\bibinfo  {journal} {arXiv preprint
  arXiv:2106.09819}\ } (\bibinfo {year} {2021})}\BibitemShut {NoStop}%
\bibitem [{\citenamefont {Deng}\ and\ \citenamefont
  {Hani}(2023{\natexlab{b}})}]{deng2023long}%
  \BibitemOpen
  \bibfield  {author} {\bibinfo {author} {\bibfnamefont {Y.}~\bibnamefont
  {Deng}}\ and\ \bibinfo {author} {\bibfnamefont {Z.}~\bibnamefont {Hani}},\
  }\bibfield  {title} {\bibinfo {title} {Long time justification of wave
  turbulence theory},\ }\href@noop {} {\bibfield  {journal} {\bibinfo
  {journal} {arXiv preprint arXiv:2311.10082}\ } (\bibinfo {year}
  {2023}{\natexlab{b}})}\BibitemShut {NoStop}%
\bibitem [{\citenamefont {Deng}\ and\ \citenamefont
  {Hani}(2021)}]{deng2021propagation}%
  \BibitemOpen
  \bibfield  {author} {\bibinfo {author} {\bibfnamefont {Y.}~\bibnamefont
  {Deng}}\ and\ \bibinfo {author} {\bibfnamefont {Z.}~\bibnamefont {Hani}},\
  }\bibfield  {title} {\bibinfo {title} {Propagation of chaos and the higher
  order statistics in the wave kinetic theory},\ }\href@noop {} {\bibfield
  {journal} {\bibinfo  {journal} {arXiv preprint arXiv:2110.04565}\ } (\bibinfo
  {year} {2021})}\BibitemShut {NoStop}%
\bibitem [{\citenamefont {J.~Bardeen}(1957)}]{BCSTheory}%
  \BibitemOpen
  \bibfield  {author} {\bibinfo {author} {\bibfnamefont {J.~R.~S.}\
  \bibnamefont {J.~Bardeen}, \bibfnamefont {L.~N.~Cooper}},\ }\bibfield
  {title} {\bibinfo {title} {Microscopic theory of superconductivity},\
  }\href@noop {} {\bibfield  {journal} {\bibinfo  {journal} {Physical Review}\
  }\textbf {\bibinfo {volume} {106}} (\bibinfo {year} {1957})}\BibitemShut
  {NoStop}%
\bibitem [{\citenamefont {Zakharov}\ \emph {et~al.}(1975)\citenamefont
  {Zakharov}, \citenamefont {Lvov},\ and\ \citenamefont
  {Starobinets}}]{zakharov1975spin}%
  \BibitemOpen
  \bibfield  {author} {\bibinfo {author} {\bibfnamefont {V.~E.}\ \bibnamefont
  {Zakharov}}, \bibinfo {author} {\bibfnamefont {V.~S.}\ \bibnamefont {Lvov}},\
  and\ \bibinfo {author} {\bibfnamefont {S.~S.}\ \bibnamefont {Starobinets}},\
  }\bibfield  {title} {\bibinfo {title} {Spin-wave turbulence beyond the
  parametric excitation threshold},\ }\href@noop {} {\bibfield  {journal}
  {\bibinfo  {journal} {Physics-Uspekhi}\ }\textbf {\bibinfo {volume} {17}},\
  \bibinfo {pages} {896} (\bibinfo {year} {1975})}\BibitemShut {NoStop}%
\bibitem [{\citenamefont {Lvov}(2012)}]{l2012wave}%
  \BibitemOpen
  \bibfield  {author} {\bibinfo {author} {\bibfnamefont {V.~S.}\ \bibnamefont
  {Lvov}},\ }\href@noop {} {\emph {\bibinfo {title} {Wave turbulence under
  parametric excitation: applications to magnets}}}\ (\bibinfo  {publisher}
  {Springer Science \& Business Media},\ \bibinfo {year} {2012})\BibitemShut
  {NoStop}%
\bibitem [{\citenamefont {Zaleski}\ \emph {et~al.}(2020)\citenamefont
  {Zaleski}, \citenamefont {Onorato},\ and\ \citenamefont
  {Lvov}}]{zaleski2020anomalous}%
  \BibitemOpen
  \bibfield  {author} {\bibinfo {author} {\bibfnamefont {J.}~\bibnamefont
  {Zaleski}}, \bibinfo {author} {\bibfnamefont {M.}~\bibnamefont {Onorato}},\
  and\ \bibinfo {author} {\bibfnamefont {Y.~V.}\ \bibnamefont {Lvov}},\
  }\bibfield  {title} {\bibinfo {title} {Anomalous correlators in nonlinear
  dispersive wave systems},\ }\href@noop {} {\bibfield  {journal} {\bibinfo
  {journal} {Physical Review X}\ }\textbf {\bibinfo {volume} {10}},\ \bibinfo
  {pages} {021043} (\bibinfo {year} {2020})}\BibitemShut {NoStop}%
\bibitem [{\citenamefont {Zakharov}\ \emph {et~al.}(2004)\citenamefont
  {Zakharov}, \citenamefont {Dias},\ and\ \citenamefont {Pushkarev}}]{MMT2}%
  \BibitemOpen
  \bibfield  {author} {\bibinfo {author} {\bibfnamefont {V.}~\bibnamefont
  {Zakharov}}, \bibinfo {author} {\bibfnamefont {F.}~\bibnamefont {Dias}},\
  and\ \bibinfo {author} {\bibfnamefont {A.}~\bibnamefont {Pushkarev}},\
  }\bibfield  {title} {\bibinfo {title} {One-dimensional wave turbulence},\
  }\href@noop {} {\bibfield  {journal} {\bibinfo  {journal} {Physics Reports}\
  }\textbf {\bibinfo {volume} {398}},\ \bibinfo {pages} {1} (\bibinfo {year}
  {2004})}\BibitemShut {NoStop}%
\bibitem [{\citenamefont {Deng}\ \emph {et~al.}(2022)\citenamefont {Deng},
  \citenamefont {Ionescu},\ and\ \citenamefont {Pusateri}}]{deng2022wave}%
  \BibitemOpen
  \bibfield  {author} {\bibinfo {author} {\bibfnamefont {Y.}~\bibnamefont
  {Deng}}, \bibinfo {author} {\bibfnamefont {A.~D.}\ \bibnamefont {Ionescu}},\
  and\ \bibinfo {author} {\bibfnamefont {F.}~\bibnamefont {Pusateri}},\
  }\bibfield  {title} {\bibinfo {title} {On the wave turbulence theory of 2d
  gravity waves, i: deterministic energy estimates},\ }\href@noop {} {\bibfield
   {journal} {\bibinfo  {journal} {Communications on Pure and Applied
  Mathematics}\ } (\bibinfo {year} {2022})}\BibitemShut {NoStop}%
\bibitem [{\citenamefont {Berti}\ \emph {et~al.}(2024)\citenamefont {Berti},
  \citenamefont {Maspero},\ and\ \citenamefont
  {Murgante}}]{berti2024hamiltonian}%
  \BibitemOpen
  \bibfield  {author} {\bibinfo {author} {\bibfnamefont {M.}~\bibnamefont
  {Berti}}, \bibinfo {author} {\bibfnamefont {A.}~\bibnamefont {Maspero}},\
  and\ \bibinfo {author} {\bibfnamefont {F.}~\bibnamefont {Murgante}},\
  }\bibfield  {title} {\bibinfo {title} {Hamiltonian birkhoff normal form for
  gravity-capillary water waves with constant vorticity: almost global
  existence},\ }\href@noop {} {\bibfield  {journal} {\bibinfo  {journal}
  {Annals of PDE}\ }\textbf {\bibinfo {volume} {10}},\ \bibinfo {pages} {22}
  (\bibinfo {year} {2024})}\BibitemShut {NoStop}%
\bibitem [{\citenamefont {Baldi}\ \emph {et~al.}(2018)\citenamefont {Baldi},
  \citenamefont {Berti}, \citenamefont {Haus},\ and\ \citenamefont
  {Montalto}}]{baldi2018time}%
  \BibitemOpen
  \bibfield  {author} {\bibinfo {author} {\bibfnamefont {P.}~\bibnamefont
  {Baldi}}, \bibinfo {author} {\bibfnamefont {M.}~\bibnamefont {Berti}},
  \bibinfo {author} {\bibfnamefont {E.}~\bibnamefont {Haus}},\ and\ \bibinfo
  {author} {\bibfnamefont {R.}~\bibnamefont {Montalto}},\ }\bibfield  {title}
  {\bibinfo {title} {Time quasi-periodic gravity water waves in finite depth},\
  }\href@noop {} {\bibfield  {journal} {\bibinfo  {journal} {Inventiones
  mathematicae}\ }\textbf {\bibinfo {volume} {214}},\ \bibinfo {pages} {739}
  (\bibinfo {year} {2018})}\BibitemShut {NoStop}%
\bibitem [{\citenamefont {Korotkevich}\ \emph {et~al.}(2024)\citenamefont
  {Korotkevich}, \citenamefont {Nazarenko}, \citenamefont {Pan},\ and\
  \citenamefont {Shatah}}]{korotkevich2024non}%
  \BibitemOpen
  \bibfield  {author} {\bibinfo {author} {\bibfnamefont {A.~O.}\ \bibnamefont
  {Korotkevich}}, \bibinfo {author} {\bibfnamefont {S.~V.}\ \bibnamefont
  {Nazarenko}}, \bibinfo {author} {\bibfnamefont {Y.}~\bibnamefont {Pan}},\
  and\ \bibinfo {author} {\bibfnamefont {J.}~\bibnamefont {Shatah}},\
  }\bibfield  {title} {\bibinfo {title} {Non-local gravity wave turbulence in
  presence of condensate},\ }\href@noop {} {\bibfield  {journal} {\bibinfo
  {journal} {Journal of Fluid Mechanics}\ }\textbf {\bibinfo {volume} {992}},\
  \bibinfo {pages} {A1} (\bibinfo {year} {2024})}\BibitemShut {NoStop}%
\bibitem [{\citenamefont {Nazarenko}(2011{\natexlab{b}})}]{nazarenko2011wave}%
  \BibitemOpen
  \bibfield  {author} {\bibinfo {author} {\bibfnamefont {S.}~\bibnamefont
  {Nazarenko}},\ }\href@noop {} {\emph {\bibinfo {title} {Wave turbulence}}},\
  Vol.\ \bibinfo {volume} {825}\ (\bibinfo  {publisher} {Springer Science \&
  Business Media},\ \bibinfo {year} {2011})\BibitemShut {NoStop}%
\bibitem [{\citenamefont {Buckmaster}\ \emph {et~al.}(2021)\citenamefont
  {Buckmaster}, \citenamefont {Germain}, \citenamefont {Hani},\ and\
  \citenamefont {Shatah}}]{buckmaster2021onset}%
  \BibitemOpen
  \bibfield  {author} {\bibinfo {author} {\bibfnamefont {T.}~\bibnamefont
  {Buckmaster}}, \bibinfo {author} {\bibfnamefont {P.}~\bibnamefont {Germain}},
  \bibinfo {author} {\bibfnamefont {Z.}~\bibnamefont {Hani}},\ and\ \bibinfo
  {author} {\bibfnamefont {J.}~\bibnamefont {Shatah}},\ }\bibfield  {title}
  {\bibinfo {title} {Onset of the wave turbulence description of the longtime
  behavior of the nonlinear schr{\"o}dinger equation},\ }\href@noop {}
  {\bibfield  {journal} {\bibinfo  {journal} {Inventiones mathematicae}\
  }\textbf {\bibinfo {volume} {225}},\ \bibinfo {pages} {787} (\bibinfo {year}
  {2021})}\BibitemShut {NoStop}%
\bibitem [{\citenamefont {V.S.Lvov}(1994)}]{LVOVBOOK}%
  \BibitemOpen
  \bibfield  {author} {\bibinfo {author} {\bibnamefont {V.S.Lvov}},\
  }\href@noop {} {\emph {\bibinfo {title} {Wave Turbulence Under Parametric
  Excitations, Applications to Magnets}}}\ (\bibinfo  {publisher}
  {Springer-Verlag},\ \bibinfo {year} {1994})\BibitemShut {NoStop}%
\end{thebibliography}
\end{document}